\documentclass[11pt,tightenlines,aps,prd,superscriptaddress,nofootinbib,longbibliography,notitlepage]{revtex4-1}
\pdfoutput=1

\usepackage{amsmath,amssymb,bm,mathtools}

\usepackage{graphicx}
\usepackage{amsfonts}
\usepackage{amssymb}
\usepackage{amsbsy}
\usepackage{amsmath}
\usepackage{latexsym}
\usepackage{bm}
\usepackage{color}
\usepackage{hyperref}
\usepackage[margin=2cm]{geometry}
\usepackage{comment}
\usepackage[normalem]{ulem}
\usepackage[makeroom]{cancel}
\usepackage{enumitem}

\usepackage{stackengine}
\stackMath
\newcommand\tenq[2][1]{%
 \def\useanchorwidth{T}%
  \ifnum#1>1%
    \stackon[0pt]{\tenq[\numexpr#1-1\relax]{#2}}{\scriptscriptstyle\sim}%
  \else%
    \stackon[1pt]{#2}{\scriptscriptstyle\sim}%
  \fi%
}

\newcommand{\de}{\mbox{d}}
\newcommand{\be}{\begin{equation}}
\newcommand{\ee}{\end{equation}}
\newcommand{\pha}{\phantom{a}}

\newcommand{\pa}{\partial}

\newcommand{\bea}{\begin{eqnarray}}
\newcommand{\eea}{\end{eqnarray}}

\newcommand{\lf}{\left}
\newcommand{\rg}{\right}

\newcommand{\rnew}{r_{\scalebox{0.65}{0}}}

\numberwithin{equation}{section}

\begin{document}

\title{\Large Generalized boundary conditions in closed cosmologies} 

\author{David Brizuela}
\email{david.brizuela@ehu.eus}
\affiliation{Department of Physics and EHU Quantum Center, University of the Basque Country UPV/EHU, Barrio Sarriena s/n, 48940 Leioa, Spain}

\author{Marco de Cesare}
\email{marco.decesare@na.infn.it}
\affiliation{Department of Physics and EHU Quantum Center, University of the Basque Country UPV/EHU, Barrio Sarriena s/n, 48940 Leioa, Spain}
\affiliation{Dipartimento di Fisica ``Ettore Pancini'', Universit{\`a} di Napoli Federico II, 80125 Napoli, Italy}
\affiliation{INFN, Sezione di Napoli, 80125 Napoli, Italy}

\begin{abstract}
Considering a generalization of the Gibbons-Hawking-York covariant boundary action
that depends on both the extrinsic and the intrinsic geometry of the boundary, we derive boundary conditions for the cosmological background and tensor perturbations in a closed universe with space-like boundaries.
We also give a general method to reconstruct the covariant boundary action starting from a given set of boundary conditions for the cosmological background. These results may be of special relevance in the context of the path-integral formulation of quantum cosmology, where boundary terms contain essential physical
information of the system.
\end{abstract}

\maketitle

\nopagebreak

\section{Introduction}

It is well known that the Einstein-Hilbert action needs to be supplemented by an appropriate boundary action in order to be functionally differentiable~\cite{Padmanabhan:2014lwa}. This ensures that the variation of the total gravitational action gives the Einstein field equations in the bulk, in combination with a suitable set of boundary conditions.
The best known example in general relativity is the Gibbons-Hawking-York (GHY) boundary action~\cite{Gibbons:1976ue,York:1986lje}, which  is defined for space-like and time-like boundaries as the integral of the trace of the extrinsic curvature, and gives Dirichlet boundary conditions for the metric on the boundary. However, other boundary conditions are possible (e.g., Neumann, Robin, and York's mixed boundary conditions) and may be better suited
in certain physical scenarios \cite{Krishnan:2016mcj,Krishnan:2017bte,York:1986lje,Odak:2021axr,Olea:2006vd}.
Moreover, the quasi-local charges of a bounded spacetime region are known to depend on boundary conditions \cite{Brown:1992br,Odak:2021axr}. Therefore, even if the equations of motion in the bulk do not depend
on these boundary terms, the latter encode key physical information of the system.

In particular, boundary conditions on space-like boundaries are of great importance in quantum cosmology, as they define the initial and final configurations of the spatial geometry that enter in the definition of the path integral. The best known examples of boundary conditions in quantum cosmology are the ``no-boundary proposal''
\footnote{Note that, in the original Euclidean formulation of the ``no-boundary proposal'', there is only one boundary on which the argument of the wave function is specified.}~\cite{Hartle:1983ai}
 and the ``tunneling proposal'' \cite{Vilenkin:1982de,Vilenkin:2018dch}. Also the functional form of the boundary action is of central importance in this context, since it contributes to the path integral and may even lead to dramatic changes in its behavior. In fact, in Ref.~\cite{DiTucci:2019dji} it was shown that, by imposing Robin boundary conditions in the Lorentzian path integral (as defined rigorously in Ref.~\cite{Feldbrugge:2017kzv} using Picard-Lefschetz theory), one obtains a stable saddle point geometry. This is in stark contrast with the results obtained by imposing Dirichlet boundary conditions, which predict unstable perturbations on the final spatial geometry \cite{Feldbrugge:2017fcc}. Different possibilities for boundary conditions in minisuperspace path integrals have been further examined in Refs.~\cite{DiTucci:2019bui,Rajeev:2021xit}. For alternative proposals that try to construct
 a framework that leads to damped perturbations on the final surface,
 see also Refs.~\cite{DiazDorronsoro:2018wro,Janssen:2019sex}.

In this paper we consider a more general boundary action defined on the space-like boundary of a closed universe, which involves an arbitrary function of both the extrinsic and intrinsic curvatures. This allows us to reproduce all known examples of boundary conditions that have been considered in Lorentzian quantum cosmology. Moreover, since we start from a covariant boundary action, this approach enables us to derive the boundary conditions for cosmological perturbations from the same action principle as the background.
In particular, we explicitly carry out the derivation of boundary conditions for the cosmological background and tensor perturbations, and our results are valid for any matter fields that do not source tensor modes (such as a scalar field or a perfect fluid). Furthermore, we also give a general method to reconstruct the precise
functional form of the covariant boundary action, given a set of boundary conditions for the cosmological background. In particular, this can be used to check whether any given set of boundary conditions prescribed for the cosmological background and perturbations can be seen as arising from a given covariant boundary action.

The remainder of the paper is organized as follows. In Sec.~\ref{Sec:Action} we give the boundary action and derive the corresponding boundary conditions for a general spacetime with compact space-like boundaries.
In Sec.~\ref{Sec:Background} we obtain the boundary conditions for the cosmological background, assuming the Friedmann-Lema{\^i}tre-Robertson-Walker (FLRW) geometry for a closed universe. In Sec.~\ref{sec:tensorperturbations}
we present the
total action for the tensor perturbations, whose variation leads to the corresponding equations of motion
and boundary conditions. The reconstruction method is given in Sec.~\ref{Sec:Reconstruction}. Finally, our results are reviewed in Sec.~\ref{Sec:Conclusion}.

{\bf Conventions.} We assume the metric signature $(-+++)$ and work in units where $c=1$. The gravitational
coupling constant is denoted as $\kappa=8\pi G$. Abstract indices for spacetime tensors are in Latin, while Greek letters are used for three-dimensional spatial tensors.

\section{Generalized boundary action}\label{Sec:Action}

We assume a globally hyperbolic spacetime $(M, g_{ab})$ foliated by
compact space-like Cauchy surfaces $M^3$,
i.e., $M\cong M^3 \times \mathbb{R}$. Using the usual $3+1$
decomposition, we introduce the spatial metric $q_{ab}\equiv g_{ab}+n_a n_b$, where $n_a$ is the normal one-form to the leaves of the foliation. The extrinsic curvature is defined as
the Lie dragging of the spatial metric $K_{ab}\equiv\frac{1}{2}\pounds_n q_{ab}$,
and its trace is denoted as~$K$. The Levi-Civita connection associated with $q_{ab}$ is denoted as $D_c$, the spatial Ricci tensor is $\mathcal{R}_{ab}$, and the spatial curvature scalar is $\mathcal{R}$.

Let us now consider a compact spacetime region $\Omega\subset M$ with space-like boundary $\pa \Omega= \Sigma_{1} \cup \Sigma_{2}$, where  $\Sigma_{1}, \Sigma_{2}\cong M^3$. The sets $(\Sigma_{1}, q^{\rm i}_{ab},K_{ab}^{\rm i})$ and $(\Sigma_{2}, q^{\rm f}_{ab},K_{ab}^{\rm f})$ can thus
be respectively regarded as the initial and final states of the geometry of a closed universe.
We also include matter fields, schematically denoted as~$\psi$; for simplicity,
we will assume these to be minimally coupled to gravity.
The action of the system is then given by
\be\label{Eq:Action}
S[g_{ab},\psi]=\frac{1}{2\kappa}\int_{\Omega} \de^4x \sqrt{-g}\, R+\frac{1}{\kappa}\int_{\Sigma_{1}} \de^3x \sqrt{q^{\rm i}}\, F_1(K^{\rm i},\mathcal{R}^{\rm i})+\frac{1}{\kappa}\int_{\Sigma_{2}} \de^3x \sqrt{q^{\rm f}}\, F_2(K^{\rm f},\mathcal{R}^{\rm f})+S_{\rm m}[g_{ab},\psi],
\ee
where the first term is the bulk Einstein-Hilbert action, supplemented
by general boundary terms (the second and third terms), while the last term corresponds to the matter action.
 The labels ${\rm i}$ and ${\rm f}$ denote
geometric objects corresponding to the initial $\Sigma_{1}$ and final $\Sigma_{2}$
space-like leaves.
The functions $F_1$ and $F_2$ are completely general, and they may be different.
Their arguments are the trace of the
extrinsic curvature $K$ and the spatial Ricci scalar ${\cal R}$, so
they depend both on the intrinsic and extrinsic geometries of the space-like boundaries. Note that in the particular case $F_1=-F_2=K$ we recover the GHY boundary action \cite{York:1986lje}, while for constant $F_1$ and $F_2$ we obtain the boundary action considered in Ref.~\cite{Krishnan:2017bte}.
In fact, higher-order curvature invariants (such as, for instance,
${\cal R}^{ab}{\cal R}_{ab}$ or $K^{ab}{\cal R}_{ab}$) may also be included in the functional dependence of $F$. However, at the level of the homogeneous and isotropic
background model we will study, such generalizations are completely equivalent to $F=F(K,\mathcal{R})$. Therefore, for simplicity,
in the present work we will confine our attention to the action~\eqref{Eq:Action}.
Also, even if in this paper we focus specifically on space-like boundaries, which are relevant for cosmological applications, generalizations of the action \eqref{Eq:Action} with time-like boundaries are straightforward.

Varying the action \eqref{Eq:Action} we obtain
\begin{align}
\delta S &=\frac{1}{2\kappa}\int_{\Omega} \de^4x \sqrt{-g}\, \lf(G_{ab}-\kappa T_{ab}\rg)\delta g^{ab}+\int_{\Sigma_{1}} \de^3x \sqrt{q^{\rm i}}\, \lf(\frac{1}{\kappa} \mathcal{B}_{\rm i,grav}+\mathcal{B}_{\rm i,matter}\rg)
\nonumber\\\label{Eq:Variation}
&+ \int_{\Sigma_{2}} \de^3x \sqrt{q^{\rm f}}\, \lf(\frac{1}{\kappa} \mathcal{B}_{\rm f,grav}+\mathcal{B}_{\rm f,matter}\rg),
\end{align}
where $T_{ab}\equiv -\frac{2}{\sqrt{-g}}\frac{\delta S_{\rm m}[\psi]}{\delta g^{ab}}$ is
the stress-energy tensor, and we have defined the boundary terms,
\be\label{EQ:BC1}
\begin{split}
\mathcal{B}_{\rm i, grav}&\equiv\left(-1+\frac{\pa F_1}{\pa K}\rg)q^{ab}_{\rm i}\delta K_{ab}^{\rm i}\\
&+\lf[\frac{1}{2}F_1 q^{ab}_{\rm i}+\lf(\frac{1}{2}- \frac{\pa F_1}{\pa K}\rg) K^{ab}_{\rm i}-\frac{\pa F_1}{\pa \mathcal{R}} \mathcal{R}_{\rm i}^{ab} + D^a D^b \frac{\pa F_1}{\pa \mathcal{R}}  +q^{ab}_{\rm i} D^c D_c \frac{\pa F_1}{\pa \mathcal{R}} \rg]\delta q^{\rm i}_{ab}=0,
\end{split}
\ee
\be\label{EQ:BC2}
\begin{split}
\mathcal{B}_{\rm f, grav}&\equiv\left(1+\frac{\pa F_2}{\pa K}\rg)q^{ab}_{\rm f}\delta K_{ab}^{\rm f}\\
&+\lf[\frac{1}{2}F_2 q^{ab}_{\rm f}-\lf(\frac{1}{2}+ \frac{\pa F_2}{\pa K}\rg) K^{ab}_{\rm f}-\frac{\pa F_2}{\pa \mathcal{R}} \mathcal{R}_{\rm f}^{ab} + D^a D^b \frac{\pa F_2}{\pa \mathcal{R}}  +q^{ab}_{\rm f} D^c D_c \frac{\pa F_2}{\pa \mathcal{R}} \rg]\delta q^{\rm f}_{ab}=0,
\end{split}
\ee
which must be vanishing for the variation of the action to be well defined.
Note that the quantities $\mathcal{B}_{\rm i, grav}$ and $\mathcal{B}_{\rm f, grav}$ also include the boundary contribution arising from the variation of the Einstein-Hilbert action\footnote{The different signs in front of the first term in Eqs.~\eqref{EQ:BC1} and \eqref{EQ:BC2} are due to the orientation of the space-like boundaries, see, e.g., Ref.~\cite{York:1986lje}.}. The matter contributions to the boundary terms must also vanish,
and the equations $\mathcal{B}_{\rm i,matter}=0$ and $\mathcal{B}_{\rm f,matter}=0$ thus
represent the matter boundary conditions. Their precise form depends on the matter type, but
we will not be concerned with these terms and leave the matter content completely general.
We stress that the matter action does not contribute to the boundary conditions for the gravitational field \eqref{EQ:BC1}--\eqref{EQ:BC2}, due to the fact that we have assumed matter fields to be minimally coupled.
This would no longer be the case for nonminimal couplings (for instance, in the presence of dilaton couplings).
Finally, we would like to point out that there are no corner terms in the variation~\eqref{Eq:Variation} since we are assuming that $\pa \Omega$
has no boundary.

The standard GHY term leads to Dirichlet boundary conditions for the spatial metric $\delta q_{ab}^{\rm i}=\delta q_{ab}^{\rm f}=0$, as can be straightforwardly checked by inserting $F_1=-F_2=K$ in \eqref{EQ:BC1}--\eqref{EQ:BC2}.
However,
our more general boundary conditions \eqref{EQ:BC1}--\eqref{EQ:BC2} involve both the intrinsic and the extrinsic geometry of the boundaries. For instance,
we can also obtain the boundary conditions of Ref.~\cite{Krishnan:2017bte}
by considering that either $F_{1}$ or $F_{2}$ are constant. For instance, in the case $F_{2}=constant$, from Eq.~\eqref{EQ:BC2},  we find the boundary condition
\be
\mathcal{B}_{\rm f,grav}=q^{ab}_{\rm f}\delta K_{ab}^{\rm f}+\frac{1}{2}\lf(F_2 q^{ab}_{\rm f}- K^{ab}_{\rm f}  \rg)\delta q^{\rm f}_{ab}=0~.
\ee

Before closing this section, we remark that the boundary conditions \eqref{EQ:BC1}--\eqref{EQ:BC2} may also be
expressed in terms of the canonical Arnowitt-Deser-Misner momentum, defined as $\pi^{ab}=\frac{\sqrt{q}}{2\kappa}\lf(K^{ab}-K q^{ab} \rg) $. In the canonical formulation of general relativity, $\pi^{ab}$ would be the natural variable to consider instead of $K^{ab}$. The translation of our results from the $(q_{ab},K^{ab})$ variables to $(q_{ab},\pi^{ab})$ is straightforward.

\section{Boundary conditions for the cosmological background}\label{Sec:Background}

Let us now assume the closed FLRW geometry for the cosmological background. Thus, in this case, the
compact spatial manifold $M^3$ is the three-sphere. In a generic time gauge the metric reads
\be
\de s^2= - N^2(t) \de t^2 + \rnew^2 a^2(t) \lf(\de \chi^2+\sin^2\chi\, \lf(\de\theta^2+ \sin^2\theta\, \de\phi^2\rg)\rg)~,
\ee
where $a$ is the scale factor, $N$ is the lapse, and $\rnew$ is the radius of the three-sphere. The Hubble rate is defined as $H\equiv \dot{a}/ (N a)$, with the dot
denoting a derivative with respect to $t$. The spatial curvature is
given by $\mathcal{R}=6/(\rnew a)^2$, and the trace of the extrinsic curvature
is $K=3H$. Therefore, the function $F_j=F_j(K,{\cal R})$, for $j=1,2$,  can be regarded now as a function
of the scale factor and the Hubble rate, i.e., $F_j=F_j(a,H)$.

Excluding the matter contribution $S_m$ and integrating by parts to remove second-order derivatives, for this metric, the action \eqref{Eq:Action} reduces to
\be\label{eq_actionFLRW}
S[a]=\frac{3}{\kappa}\int_{\Omega} d^4x \sqrt{\gamma}\,a N\left(\frac{1}{\rnew^2}-\frac{\dot{a}^2}{N^2} \right)
+\sum_{j=1}^2\frac{1}{\kappa}\int_{\Sigma_{j}} d^3 x \sqrt{\gamma}\,a^3\left(F_j+3(-1)^jH \right)~,
\ee
with $\gamma$ being the determinant of the metric of the three-sphere, which will be explicitly defined below.
In turn,
the boundary conditions \eqref{EQ:BC1}--\eqref{EQ:BC2} take now
the simpler form
\be\label{EQ:Background_BC}
\lf((-1)^j+\frac{1}{3}\frac{\pa F_j}{\pa H}\rg) \delta H +\lf(F_j+\frac{1}{3}a\frac{\pa F_j}{\pa a}+(-1)^j H \rg)\frac{\delta a}{a} =0~.
\ee
Note that the variables $(a, H)$ are evaluated on the boundaries
$\Sigma_{1}$ and $\Sigma_{2}$ for $j=1$ and $j=2$, respectively. In principle, all these quantities should carry a $j$ label to denote
the boundary where they are evaluated.
However, in order to make the notation lighter, we will omit it, except for the function $F_j$, since the initial and final boundary Lagrangians are not necessarily the same.

The boundary conditions \eqref{EQ:Background_BC} are manifestly gauge invariant under time reparametrizations (in that they do not involve the lapse)
and represent a generalization of the standard Dirichlet boundary conditions $(\delta a=0)$ obtained from the GHY boundary action for cosmology. In Sec.~\ref{Sec:Reconstruction} we will give a general procedure to reconstruct the functional form of $F_j$ from any given boundary conditions obeyed by the cosmological background.

\section{Tensor perturbations}\label{sec:tensorperturbations}

If one considers tensor perturbations $h_{\alpha\beta}$
propagating on the background given by the FLRW universe presented above, up
to linear order in $h_{\alpha\beta}$, the metric takes the form
\be\label{Eq:PerturbedMetric}
\de s^2= - N^2(t) \de t^2 +a^2(t) (\gamma_{\alpha\beta}(x^\gamma)+h_{\alpha\beta}(t,x^\gamma))\de x^\alpha \de x^\beta ~,
\ee
where
\be
\gamma_{\alpha\beta}\de x^\alpha \de x^\beta=\rnew^2 \left(\de \chi^2+\sin^2\chi\,\left( \de\theta^2+\sin^2\theta\,\de\phi^2\right)\right)~,
\ee
denotes the metric on the three-sphere. Note that, at the level of the background geometry, the spatial metric $q_{\alpha\beta}$ and $\gamma_{\alpha\beta}$ are conformally related, $q_{\alpha\beta}=a^2\, \gamma_{\alpha\beta}$. 
In the following, as it is usual in cosmological perturbation theory,
we will use $\gamma_{\alpha\beta}$ to raise the spatial (Greek) indices.
By definition, tensor perturbations are transverse and traceless,
\be
\gamma^{\alpha \beta}h_{\alpha\beta}=0~,\quad h_{\alpha\beta}^{\pha\pha|\beta}=0~,
\ee
where the vertical bar denotes the covariant derivative compatible with the metric
$\gamma_{\alpha\beta}$, i.e., $\gamma_{\alpha\beta|\lambda}=0$.

\subsection{Action for tensor perturbations}

The boundary conditions obeyed by the tensor perturbations are completely fixed
by the covariant action~\eqref{Eq:Action}. In order to obtain such conditions, 
we will first derive the total action for the perturbations that, upon variation, will provide
the sought-for boundary conditions.
Our aim in this section is thus to compute the total action for the perturbations, including both the bulk
and boundary contributions. This will be given by the second-order
variation of the background action. Therefore, we first need to expand all relevant
quantities that characterize the intrinsic and extrinsic geometry up to second order (that is,
quadratic terms) in the perturbations.

The spatial metric and extrinsic curvature obtained from the perturbed metric \eqref{Eq:PerturbedMetric} read as
\be\label{Eq:PerturbedSpatialGeom}
q_{\alpha\beta}=a^2 (\gamma_{\alpha\beta}+h_{\alpha\beta}) ~, \quad K_{\alpha\beta}=a^2  \lf(H (\gamma_{\alpha \beta}+ h_{\alpha \beta})+\frac{1}{2N}\dot{h}_{\alpha \beta}\rg)~,
\ee
respectively, which, given the metric \eqref{Eq:PerturbedMetric}, are exact expressions.  
Up to quadratic terms in the perturbations, the inverse spatial metric is given by
\be
q^{\alpha\beta}\simeq a^{-2} (\gamma^{\alpha\beta}-h^{\alpha\beta}+h^{\alpha}_{\pha\gamma} h^{\gamma\beta})~,
\ee
while for the spatial Ricci scalar we have
\be\label{Eq:SpatialRicciScalar}
{\cal R}\simeq\frac{6}{a^2 \rnew^2}-\frac{1}{a^2}\lf(\frac{1}{\rnew^2}h_{\alpha\beta}h^{\alpha\beta} - h^{\alpha\beta} h_{\alpha\beta\pha|\lambda}^{\pha\pha\;|\lambda}+\frac{1}{2}h^{\alpha\lambda|\beta} h_{\alpha\beta|\lambda}-\frac{3}{4}h^{\alpha\beta|\lambda} h_{\alpha\beta|\lambda}\rg)~,
\ee
and the volume element is
\be
\sqrt{-g}\simeq N a^3 \sqrt{\gamma}\lf(1-\frac{1}{4} h_{\alpha\beta}h^{\alpha\beta}\rg)~.
\ee

Making use of the Gauss-Codazzi relations, the four-dimensional Ricci scalar can be written in a
generic gauge as follows,
\be
R= {\cal R}+K_{ab}K^{ab}-K^2+2\nabla_a\lf( n^a \nabla_c n^c - n^c \nabla_c n^a\rg),
\ee
Note that, in this expression, one needs to use the full four-dimensional metric to raise and lower indices, and
this is the reason for using Latin indices here.
In the gauge defined by Eq.~\eqref{Eq:PerturbedMetric}, the normal vector is $n^a=\frac{1}{N}\lf(\frac{\pa}{\pa t}\rg)^a$, whose acceleration vanishes, $n^c \nabla_c n^a=0$,\footnote{This is because we have $n^c \nabla_c n^a=D_a \log N=0$, where $D_a$ denotes the Levi-Civita connection of the spatial metric $q_{ab}$ (see, e.g., Ref.~\cite{Gourgoulhon:2007ue}), and the last step follows from the fact that $N$ does not depend on the spatial coordinates in the chosen gauge.}
which simplifies the above relation.
Therefore, after integration by parts, we get that the Einstein-Hilbert actions reads,
\be\label{Eq:ExpandedActionGauss}
S_{\rm EH}=\frac{1}{2\kappa}\int_{\Omega} \de^4x \sqrt{-g} \lf({\cal R}+K_{ab}K^{ab}-K^2\rg)
+\sum_{j}\frac{(-1)^j}{\kappa} \int_{\Sigma_{j}} \de^3x \sqrt{q}\, K~,
\ee
where, as in the previous section, $j=1,2$ stands for the initial and final surface, respectively.
Substituting the expansions obtained above for the different objects that appear in this action,
and retaining only terms that are quadratic in the tensor perturbations, we obtain,
after integrating by parts once again and some straightforward manipulations,
\begin{align}
S^{(2)}_{\rm EH}[h_{ab}]&=\!\int_{\Omega} \de^4x\, \frac{\sqrt{\gamma}}{8\kappa} \, N a \lf[ \lf(\frac{a}{N}\rg)^2 \dot{h}_{\alpha\beta}\dot{h}^{\alpha\beta} - h_{\alpha\beta|\lambda}h^{\alpha\beta|\lambda} - \frac{2}{\rnew^2} h_{\alpha\beta}h^{\alpha\beta} -2a^2\lf(3H^2+\frac{2}{N}\dot{H} +\frac{1}{a^2 \rnew^2}\rg)h_{\alpha\beta}h^{\alpha\beta} \rg]
\nonumber\\\label{Eq:SecondOrderActionEH}
&+\sum_{j} \frac{(-1)^{j+1}}{2\kappa}\int_{\Sigma_{j}}\de^3x\,\sqrt{\gamma}\, a^3 \lf(  \frac{1}{N} \dot{h}_{\alpha\beta}h^{\alpha\beta} + \frac{H}{2} h_{\alpha\beta}h^{\alpha\beta} \rg)~,
\end{align}
where $\sqrt{\gamma}=r_o^3 \sin^2\chi\sin\theta$. In addition,
assuming a vanishing anisotropic stress, the second-order expansion of the matter action
leads to the contribution,
\be\label{Eq:SecondOrderActionMatter}
S^{(2)}_{\rm m}[h_{ab},\psi]=-\frac{1}{4}  \int_{\Omega} \de^4x\, \sqrt{\gamma}\, N a^3  h_{\alpha\beta}h^{\alpha\beta} \overline{p}~,
\ee
where $\overline{p}$ is the pressure.\footnote{Pressure is defined as usual: the spatial components of the matter stress-energy tensor are all equal at the background level and given by $\overline{T}^{\alpha}_{\pha \beta}=\overline{p}\, \delta^{\alpha}_{\pha \beta}$ \cite{Wald:1984rg}.} Using the background field equations, the term \eqref{Eq:SecondOrderActionMatter} exactly cancels the term that multiplies $h_{\alpha\beta}h^{\alpha\beta}$ inside the
bulk integral in Eq.~\eqref{Eq:SecondOrderActionEH}.

On the other hand,
the contribution from the boundary action \eqref{Eq:Action} can be easily computed,
\be\label{Eq:SecondOrderActionB}
S^{(2)}_{\rm B}[h_{ab}]\!=\!\sum_{j=1}^2  \int_{\Sigma_ j} \de^3x \frac{\sqrt{\gamma}}{\kappa} \lf[ \lf(-\frac{1}{4}a^3 \overline{F_j} +\frac{a}{2 \rnew^2}\overline{\frac{\pa F_j}{\pa {\cal R}}} \,\rg)h_{\alpha\beta}h^{\alpha\beta}-\frac{a}{4} h_{\alpha\beta|\lambda}h^{\alpha\beta|\lambda}\overline{\frac{\pa F_j}{\pa {\cal R}}} -\frac{a^3}{2N}\dot{h}_{\alpha\beta}h^{\alpha\beta}\overline{\frac{\pa F_j}{\pa K}}\,\rg],\!
\ee
where the overbar denotes quantities evaluated on the background.

Finally, combining Eqs.~\eqref{Eq:SecondOrderActionEH} and \eqref{Eq:SecondOrderActionB}, we obtain the bulk
and boundary action for the tensor perturbations:
\begin{align}
\label{Eq:SecondOrderActionBulk}
S^{\rm pert}_{\rm bulk}[h_{ab}] &=\frac{1}{8\kappa}\int_{\Omega} \de^4x\, \sqrt{\gamma} \, N a \lf( \lf(\frac{a}{N}\rg)^2 \dot{h}_{\alpha\beta}\dot{h}^{\alpha\beta} - h_{\alpha\beta|\lambda}h^{\alpha\beta|\lambda} - \frac{2}{r_o^2} h_{\alpha\beta}h^{\alpha\beta} \rg)~,\\\nonumber
S^{\rm pert}_{\rm boundary}[h_{ab}]&=-\frac{1}{\kappa}\sum_{j=1}^2 \int_{\Sigma_j} \de^3x \sqrt{\gamma} \lf[ \lf((-1)^{j}\frac{a^3 H}{4} +\frac{1}{4}a^3 \overline{F_j} -\frac{a}{2\rnew^2} \overline{\frac{\pa F_j}{\pa {\cal R}}} \,\rg)h_{\alpha\beta}h^{\alpha\beta}+\frac{a}{4} h_{\alpha\beta|\lambda}h^{\alpha\beta|\lambda}\overline{\frac{\pa F_j}{\pa {\cal R}}} \rg.\\
&\lf.+\frac{a^3}{2N}\lf((-1)^{j}+\overline{\frac{\pa F_j}{\pa K}}\rg)\dot{h}_{\alpha\beta}h^{\alpha\beta}  \rg]~.
\label{Eq:SecondOrderActionBoundary}
\end{align}

\subsection{Boundary conditions}

We can now take the variation of
the total action for tensor perturbations $S^{\rm pert}[h_{ab}]=S^{\rm pert}_{\rm bulk}[h_{ab}]+S^{\rm pert}_{\rm boundary}[h_{ab}]$, which leads to
the equations of motion
\be
\frac{1}{Na}\frac{\de}{\de t}\lf(\frac{a^3}{N}\dot{h}_{\alpha\beta}\rg)-h_{\alpha\beta|\lambda}^{\pha\pha\pha\pha|\lambda}-\frac{2}{\rnew^2}h_{\alpha\beta}=0~,
\ee
along with the boundary conditions
\be\label{Eq:BoundaryConditionTensorPert}
A^{\alpha\beta}\delta h_{\alpha\beta}+B^{\alpha\beta} \delta \dot{h}_{\alpha\beta}=0~,
\ee
where we have defined
\begin{align}
A^{\alpha\beta}&\equiv-\lf((-1)^{j}\frac{H}{2} +\frac{\overline{F_{j}}}{2}-\frac{1}{a^2 \rnew^2}\overline{\frac{\pa F_j}{\pa \mathcal{R}}} \rg)h^{\alpha\beta} +\frac{1}{2a^2}\overline{\frac{\pa F_j}{\pa \mathcal{R}}} h^{\alpha\beta\pha|\lambda}_{\pha\;\;|\lambda} -\frac{1}{2N}\lf(\frac{1}{2}(-1)^{j}+\overline{\frac{\pa F_j}{\pa K}}\rg)\dot{h}^{\alpha\beta}~,\\
B^{\alpha\beta}&\equiv-\frac{1}{2N}\lf((-1)^{j}+\overline{\frac{\pa F_j}{\pa K}}\rg)h^{\alpha\beta}~.
\end{align}
Recall again that $j=1,2$ stands for the initial and final surfaces, respectively, and the overbar indicates
that the corresponding magnitude should be evaluated at the background level.
Therefore, the derivatives of $F_j$ with respect to $K$ and ${\cal R}$ can also
be rewritten as derivatives with respect to $a$ and $H$.
In summary, the boundary conditions \eqref{Eq:BoundaryConditionTensorPert} obeyed by the tensor perturbations,
along with the total perturbative action, which includes both the bulk \eqref{Eq:SecondOrderActionBulk} and boundary
\eqref{Eq:SecondOrderActionBoundary} contributions, constitute the main results of this section.

\section{Reconstruction method}\label{Sec:Reconstruction}

At the end of the previous section we have derived
Eq.~\eqref{Eq:BoundaryConditionTensorPert}, which expresses the boundary conditions for tensor perturbations obtained from the action~\eqref{Eq:Action}.
As already commented above,
boundary conditions for the perturbations cannot be prescribed in an independent way
from the boundary conditions for the cosmological background: once the functions
$F_j$ are fixed, Eqs.~\eqref{EQ:Background_BC} and \eqref{Eq:BoundaryConditionTensorPert} provide the background
and linearized boundary conditions, respectively.

However, starting from certain
background boundary conditions $U(a,H)=c$, as it is done in some studies about
the path-integral approach to quantum cosmology (see, e.g., Ref. \cite{DiTucci:2019bui}),
it is unclear how to obtain the functional form of $F_j$.
In this section we will present a general
method that, given the symmetry-reduced boundary conditions $U(a,H)=c$,
will provide the functional form of $F_j$,
which will in turn unambiguously define the full covariant action principle \eqref{Eq:Action}
and the boundary conditions \eqref{Eq:BoundaryConditionTensorPert} obeyed by the tensor perturbations.
Conversely, given a set of boundary conditions for the background and tensor perturbations, it will be possible to check whether they arise from one and the same boundary action of the form~\eqref{Eq:Action}.

\subsection{General method}

Let us now consider, for
the symmetry reduced background, boundary conditions of the general form
\be\label{Eq:generalBC}
U(a,H)=c ~,
\ee
where $c$ is a constant, and its differential form $\delta U=0$ defines the variation.
Physically, the quantity $U$ corresponds to a certain combination of $a$ and $H$ that is kept fixed on the boundary when the action is varied. In particular,
all known boundary conditions in cosmology, including the familiar examples of Dirichlet $U(a,H)=a$, Neumann $U(a,H)=a^2H$, and Robin $U(a,H)=\alpha a^n+\beta a H$ boundary conditions, as well as the generalizations discussed in Ref.~\cite{DiTucci:2019bui}, can be recast in the form \eqref{Eq:generalBC}.

Our goal is thus to obtain the function $F_j$, from the condition \eqref{EQ:Background_BC}, for
a given form of the combination $U=U(a,H)$.
At this point, it turns out to be convenient to perform a change of coordinates from $(a,H)$ to $(U,\phi)$, where $\phi=\phi(a,H)$ is an arbitrary function subject to the only requirement that the Jacobian of the transformation,
\be
J\equiv \frac{\partial U}{\partial a}\frac{\pa \phi}{\pa H}- \frac{\pa U}{\pa H}\frac{\partial \phi}{\partial a}~,
\ee
is non-trivial.
Taking into account the identities between the different derivatives,
\be
\Bigg(
\begin{matrix}
\frac{\partial U}{\partial a} & \frac{\partial \phi}{\partial a}\\
\frac{\partial U}{\partial H} & \frac{\partial \phi}{\partial H}
\end{matrix}
\Bigg)
=
J
\Bigg(
\begin{matrix}
\frac{\partial H}{\partial \phi} & -\frac{\partial H}{\partial U}\\
-\frac{\partial a}{\partial \phi} & \frac{\partial a}{\partial U}
\end{matrix}
\Bigg)~,
\ee
the boundary condition \eqref{EQ:Background_BC} is rewritten as
 \begin{align}
&\left[a\left(3(-1)^j+\frac{\partial F}{\partial H}\right)\frac{\partial U}{\partial a} -\left(3(-1)^j H+3 F+a \frac{\partial F}{\partial a}\right)\frac{\partial U}{\partial H}\right]\delta\phi\;+\nonumber\\
&-\left[a\lf(3(-1)^j+ \frac{\partial F}{\partial H}\rg)\frac{\partial \phi}{\partial a} -\left(3(-1)^j H+3 F+a \frac{\partial F}{\partial a}\right)
\frac{\partial\phi}{\partial H}\right]\delta U=0~.
\end{align}
Since we have defined the variation as keeping the combination $U$ fixed, $\delta U=0$,
it is clear from this expression that the action principle will
be well defined only if the coefficient of $\delta\phi$ vanishes,
\begin{equation}
\left((-1)^j+\frac{1}{3} \frac{\partial F_j}{\partial H}\right)a\frac{\partial U}{\partial a} -\left((-1)^j H+ F_j+\frac{1}{3}a \frac{\partial F_j}{\partial a}\right)\frac{\partial U}{\partial H}=0.
\end{equation}
This provides the differential equation we were looking for, and its solution will define $F_j$
in terms of the chosen combination $U$.

In order to obtain the general solution for the above equation, one first replaces the derivatives
acting on $F_j$ to derivatives with respect
to $(U,\phi)$ by simply making use of the chain rule, which leads to
\be\label{Eq:Reconstruct}
(-1)^j\lf(a\frac{\partial U}{\partial a} -H\frac{\pa U}{\pa H}  \rg)+\frac{1}{3}J \frac{\pa F_j}{\pa \phi}=\frac{\pa U}{\pa H} F_j~.
\ee
After some straightforward manipulations, this can be recast as
\be
\frac{\pa(a^3F_j)}{\pa\phi}+3a^2(-1)^j \frac{\pa(aH)}{\pa\phi}=0~,
\ee
whose general solution is
\be\label{Eq:GeneralSolutionF}
F_j(a,H)=\frac{f\big(U(a,H)\big)}{a^3}+\frac{3(-1)^{j+1}}{a^3} \int\de\phi\; a^2\frac{\pa(aH)}{\pa\phi}~,
\ee
where $f$ is an arbitrary function. This is the most relevant result of this section and
explicitly provides the function $F_j$ in terms of the combination $U(a,H)$, which is the background
symmetry-reduced boundary condition. Once $F_j$ is obtained, the covariant boundary terms of the action
of the model are defined, which could be used for applications to nonsymmetric configurations,
and, in particular, the boundary conditions obeyed by the tensor perturbations \eqref{Eq:BoundaryConditionTensorPert} are fixed. Note that the integration function $f$ only contributes to the total action with an additive constant $f(c)$ (since, at the boundary, $U=c$). However, the specific form of the coefficients $A^{\alpha\beta}$ and $B^{\alpha\beta}$ of the perturbative boundary conditions \eqref{Eq:BoundaryConditionTensorPert} will depend both on $f(c)$
and its first-order derivative $f'(c)$.

\subsection{Examples of noteworthy boundary conditions}\label{Sec:ReconstructionExamples}

In the following we examine some examples of relevant boundary conditions that have been considered in the literature, and reconstruct the corresponding $F_j$ making use of the reconstruction method
explained above. In order not to introduce further definitions, in all the cases we will denote by $f$ the free integration function and redefine it whenever it is convenient.

\subsubsection{ Dirichlet boundary conditions}

In this case the value of the scale factor is fixed at the boundary
\be\label{Eq:BC_Dirichlet}
U(a)=a~.
\ee
Imposing $\delta a=0$ in the boundary condition \eqref{EQ:Background_BC}, it simply reads
\be
(-1)^j+\frac{1}{3}\frac{\pa F_j}{\pa H}=0~,
\ee
which can be readily solved,
\be
F_j=3(-1)^{j+1}H+f(a)~,
\ee
where $f$ is an arbitrary function of $a$. This solution can also be obtained from the general
form \eqref{Eq:GeneralSolutionF} by choosing, for instance, $\phi=H$.

The form of the boundary action can now be re-expressed in a covariant fashion as a function of the intrinsic and extrinsic curvatures of the background as
\be
F_j=(-1)^{j+1}K+f(\mathcal{R})~,
\ee
where the first term defines the standard GHY action and the function $f$ has been adequately redefined.
Replacing this form in the perturbative boundary conditions \eqref{Eq:BoundaryConditionTensorPert}, it is easy to see that
the coefficient $B^{\alpha\beta}$ is vanishing, which, in general, also implies a Dirichlet problem
for the perturbations $A^{\alpha\beta}\delta h_{\alpha\beta}=0$.

\subsubsection{Fixed Hubble rate}

We require now a fixed value of the Hubble rate at the boundary, that is,
\be
U(H)=H~.
\ee
In such a case, the boundary condition \eqref{EQ:Background_BC} takes the form
\be
F_j+\frac{1}{3}a\frac{\pa F_j}{\pa a}+(-1)^jH=0~,
\ee
with the general solution given by
\be
F_j=(-1)^{j+1}H+\frac{f(H)}{ a^{3}}~,
\ee
where $f$ is an arbitrary function of $H$. It is straightforward to get
this solution from the general form \eqref{Eq:GeneralSolutionF} by choosing, for instance, $\phi=a$.
In covariant form, after a suitable redefinition of the function $f$, it reads
\be
F_j=\frac{(-1)^{j+1}}{3}K+f(K)\mathcal{R}^{3/2}~.
\ee
Contrary to the Dirichlet conditions considered above,
this form of the background boundary action does not lead to the vanishing of neither
of the coefficients $A^{\alpha\beta}$ nor $B^{\alpha\beta}$ in \eqref{Eq:BoundaryConditionTensorPert}. Therefore,
the boundary conditions for the perturbations will have the Robin form $A^{\alpha\beta}\delta h_{\alpha\beta}+ B^{\alpha\beta}\delta \dot{h}_{\alpha\beta}=0$.

\subsubsection{Neumann boundary conditions}

Neumann boundary conditions correspond to having a fixed value for the canonical momentum of $a$.
Taking the variation of the background action \eqref{eq_actionFLRW} with respect to $\dot{a}$,
it is straightforward to see that the momentum of $a$ is proportional to $a^2H$. Hence we consider
\be
U(a,H)=a^2 H~.
\ee
The corresponding solution for $F_j$ is
\be
F_j=3(-1)^j H+ \frac{f(a^2 H)}{a^{3}}~,
\ee
with $f$ being arbitrary. Its covariant form reads
\be
F_j=(-1)^j K+ {\cal R}^{3/2} f(K/{\cal R})~.
\ee
Also in this case the boundary conditions for tensor perturbations have the general Robin form.

\subsubsection{Generalized Robin boundary conditions}

A generalization of the Robin boundary conditions for the background considered in Refs.~\cite{DiTucci:2019dji,DiTucci:2019bui} is obtained with the following choice\footnote{In the notation of Ref.~\cite{DiTucci:2019dji}, the FLRW metric is parametrized as
$\de s^2= - \frac{N^2}{q} \de t^2 + q\,\lf(\de \chi^2+\sin^2\chi\, \,\left( \de\theta^2+\sin^2\theta\,\de\phi^2\right)\rg)$, and the boundary condition on the initial slice read as $\mathcal{B}=\dot{q}_o/N+\lambda q_o=constant$. The form of $\mathcal{B}$ when expressed in terms of these variables justifies the name ``Robin boundary conditions''. In Refs.~\cite{DiTucci:2019bui,Krishnan:2017bte} a different set of ``Robin boundary conditions'' was also examined, $\mathcal{B}=\dot{q}_o/N+\lambda \sqrt{q_o}=constant$. Both choices can be obtained as particular cases of Eq.~\eqref{Eq:URobin}, with $n=2$ and $n=1$, respectively.}:
\be\label{Eq:URobin}
U(a,H)=a H+\frac{\lambda}{\rnew} a^{n} ~,
\ee
with $n$ being a positive integer and $\lambda$ a dimensionless constant.
By defining any functional form for $\phi=\phi(a,H)$, which is linearly independent to this $U(a,H)$, one can perform the integral that appears
in the general solution \eqref{Eq:GeneralSolutionF}, and obtain
\be
F_j(a,H)= (-1)^j\frac{3 \lambda}{\rnew} \frac{n}{2+n} a^{(n-1)}+\frac{1}{a^{3}} f\left(a H+\frac{\lambda}{\rnew} a^{n}\right)~,
\ee
for any function $f$.
This solution can be re-expressed in covariant fashion as
\be
F_j(\mathcal{R},K)=(-1)^j \frac{3\lambda}{\rnew^n} \frac{n}{2+n} \lf(\frac{6}{\mathcal{R}}\rg)^{(n-1)/2}+\mathcal{R}^{3/2} f\lf( K \mathcal{R}^{-1/2} + 2  \lambda\, 6^{\frac{n-3}{2}} \rnew^{-n}\mathcal{R}^{-n/2}\rg),
\ee
which leads to a problem of the general form $A^{\alpha\beta}\delta h_{\alpha\beta}+ B^{\alpha\beta}\delta \dot{h}_{\alpha\beta}=0$
for the perturbations.

\section{Conclusions}\label{Sec:Conclusion}

We have considered a generalization of the GHY surface term that provides a well-defined
variational action principle for general relativity. More precisely, assuming compact
Cauchy slices, the usual
Einstein-Hilbert bulk action \eqref{Eq:Action} has been supplemented by surface terms with a generic
dependence on the three-dimensional Ricci scalar and the extrinsic curvature of the spatial slices.
The boundary conditions \eqref{EQ:BC1}--\eqref{EQ:BC2} implied by such an action have been explicitly obtained
without any symmetry assumptions.
Assuming then a perturbative framework around a FLRW closed cosmology,
boundary conditions for the cosmological background \eqref{EQ:Background_BC}
and tensor perturbations \eqref{Eq:BoundaryConditionTensorPert} have also been obtained.
In addition, the total action, including the bulk \eqref{Eq:SecondOrderActionBulk} and boundary \eqref{Eq:SecondOrderActionBoundary} contributions, for the
tensor perturbations has been explicitly derived.

The general set of boundary conditions obtained in this paper includes as special cases all known examples of boundary conditions considered in the quantum-cosmology literature. In this context, there are several
prescriptions proposed in the literature for the boundary conditions that the (symmetry-reduced)
background objects should obey. However, it is unclear whether such conditions may arise from some covariant
action. In this sense, one of the main results of the present
paper is given by the reconstruction method presented in Sec.~\ref{Sec:Reconstruction} and, more specifically, by Eq.~\eqref{Eq:GeneralSolutionF}, which provides the explicit form of the
covariant boundary action in terms of any chosen background conditions.
Since the equations for the perturbations arise from the same action principle, the boundary
conditions obeyed by the perturbations are not independent from those obeyed by the background,
and our method allows for an exact reconstruction of their general form.
Finally,
the special cases of Dirichlet, fixed Hubble rate, Neumann, and Robin boundary conditions have been examined in detail in Sec.~\ref{Sec:ReconstructionExamples}.

\section*{Acknowledgments}

We are grateful to Jean-Luc Lehners for interesting discussions and for helpful comments on a previous draft of this paper.
This work has been supported by
the Basque Government Grant \mbox{IT1628-22} and
by the Grant PID2021-123226NB-I00 (funded by MCIN/AEI/10.13039/501100011033 and by ``ERDF A way of making Europe''). The work of MdC is supported by Ministero dell'Universit{\`a} e Ricerca (MUR) (Bando PRIN 2017, Codice Progetto: 20179ZF5K5\_006) and by INFN (Iniziative specifiche QUAGRAP and GeoSymQFT).

\bibliography{QCBCrefs}

\end{document}